# On the topology of the world exchange arrangements web

Xiang Li [1]   Yu Ying Jin [2]   Guanrong Chen [3]

1. Dept. of Automation, Shanghai Jiao Tong University, Shanghai, 200030, P.R. China
2. Dept. of World Economy, Shanghai University of Finance and Economics, Shanghai, 200433, P.R. China
3. Dept. of Electronic Engineering, City University of Hong Kong, Hong Kong SAR, P.R. China

(xli@sjtu.edu.cn , and jyyshang@mail.shufe.edu.cn, and eegchen@cityu.edu.hk)

**Abstract:** Exchange arrangements among different countries over the world are foundations of the world economy, which generally stand behind the daily economic evolution. As the first study of the world exchange arrangements web (WEAW), we built a bipartite network with countries as one type of nodes and currencies as the other, and found it to have a prominent scale-free feature with a power-law degree distribution. In a further empirical study of the currency section of the WEAW, we calculated the clustering coefficients, average nearest-neighbors degree, and average shortest distance. As an essential economic network, the WEAW is found to be a correlated disassortative network with a hierarchical structure, possessing a more prominent scale-free feature than the world trade web (WTW).

Correspondence author:

   Dr. Xiang Li
   Dept. of Automation
   Shanghai Jiao Tong University
   1954 Huashan Road
   Shanghai 200030
   P.R. China

# I. Introduction

Generally, the topology of a network is mathematically described by a graph. Two recent discoveries of scale-free and small-world network models [1-2] have been able to successfully describe the topological properties of many real-life complex networks, such as the Internet, WWW, metabolic networks, language web, and many social and biological networks [1-8]. Some common features in these complex networks include power-law degree distributions (the scale-free feature), and large clustering coefficients with small average path lengths (the small-world feature), which cannot be well explained by the ever-dominating and now-classical Erdös-Rényi (ER) random graph theory [9-10].

The economic globalization promotes different countries' economies to be integrated together in terms of trade and capital flow [11-12]. The countries' economic activities are considered as being organized within a network structure, which does have a significant influence on the observed economic collective behaviors [13-15]. The viewpoints of complex networks are of interest in studying economic networks, to uncover the structural characteristics of the networks built on competitive economic systems. Serrano and Boguñá [16] first explored the topology of the world trade web (WTW), announcing that the WTW has scale-free and small-world features in a hierarchical structure, similar to the Internet and WWW. Li, Jin and Chen [14] then statistically rebuilt the WTW, and studied the economic-cycle synchronous phenomenon on this scale-free trade web. Other studies on Japanese business network [17], urban economic network [18], and stock price network [19] have also found power-law degree distributions.

Exchange rate is an important factor that can influence a country's conditions to participate international economic cooperation and competition. Exchange arrangement is the policy that shapes the exchange rate behaviors. It is argued, for example, that the financial crises in East Asia in 1997 was an outcome of the fixed exchange arrangements employed by those Eastern Asian countries. It was then suggested by many economists that a country should have more flexible exchange arrangements in order to avoid such crises [20]. Every country has its own exchange arrangement. Therefore, there exists a world exchange arrangements web (WEAW), which provides the channels to transfer a country's economic change to other countries. The web makes countries more interdependent than ever. For instance, the strengthening of USD versus Japanese Yen can give a kind of sluggish effect to East Asian economies through the exchange arrangements among them [20]. As the foundation of the world economy, the topological structure of the WEAW unavoidably and in effect significantly influences the collective economic behaviors such as economic dynamics spreading and crises propagation. It should be noted that the topology of the WEAW does not have significant change in a long period since the collapse of the Bretton Woods System, because any country would not like to adjust its exchange-rate regime frequently. Therefore, to the specific interest of this paper, we attempt to uncover the topological properties of the WEAW, leaving a further study of its complex dynamical features to the near future.

The paper is organized as follows: In Sec. II, the classification of various exchange arrangements is introduced, and the WEAW is defined and constructed. Section III investigates and visualizes the topological characteristics of the WEAW, such as the degree distribution, clustering coefficient, average path length, average nearest-neighbors degree, etc. Section IV concludes the paper with some discussions.

## II. The WEAW

Exchange rate is the price of a foreign currency in units of local currency or, conversely, the price of a local currency in units of foreign currency. Every country has its own exchange arrangement to determine the exchange rates of its local currency.

In the statistics of International Monetary Fund annual report, there are totally eight categories of exchange arrangements [21]: (1) exchange arrangement with no separate legal tender, (2) currency board arrangement, (3) conventional pegged arrangement, (4) pegged exchange rate within horizontal bands, (5) crawling peg, (6) crawling band, (7) managed floating with no pre-announced path for the exchange rate, (8) independently floating. Under independently floating exchange arrangement, it is generally believed that the exchange rate is determined by the foreign exchange market, that means the government authority does not intervene the exchange- rate behaviors in the foreign exchange market. Under the other seven arrangements, local currency is generally pegged to one or more foreign currencies in different degrees, which means the government authority intends to maintain the stability of the exchange rates of its local currency versus some foreign currencies to some preferred degree. IMF [21] has collected official exchange arrangements of 152 countries, where only 50 countries officially stated that they employ independently floating exchange arrangements (Table 1).

As defined in Table 1, the allowable flexibility degree of exchange rate under different arrangements are increased from exchange arrangement with no separate legal tender to independently floating. For simplicity, the eight exchange arrangements are generally divided into two basic groups [20, 22]: Pegged arrangement and floating arrangement. Pegged arrangement implies an explicit or understood commitment undertaken by the policy authorities to limit the extent of fluctuation of the exchange rate. Quite a broad range of regimes share this general characteristic, from exchange arrangement with no separate legal tender, to currency board arrangement, to conventional pegged arrangement, to pegged exchange rate within horizontal bands, to crawling peg, to crawling band, and to managed floating with no pre-announced path for the exchange rate. Floating arrangement refers to independently floating [20]. A recent study [22] reveals the exchange arrangements of developing countries in the period of 1990-1999, and totally 158 developing countries and regions employed the pegged exchange arrangement, in which some countries officially adopt the floating arrangement.

**Table 1: IMF classification of eight categories of exchange-rate arrangements**

| Exchange Arrangement Classification | IMF Definition | Countries with this feature | Flexibility |
|---|---|---|---|
| Exchange arrangement with no separate legal tender | **No own single legal tender**, Use other country's currency, or share the currency of a currency union as a member | 38 | Small |
| Currency board arrangement | **Explicit legislative commitment** to a **central rate** | 8 | |
| Conventional pegged arrangement | Peg, **central rate**, **narrow fluctuation margin** at $\pm 1\%$ | 45 | |
| Pegged exchange rate within horizontal bands | **Central rate**, **wider fluctuation margin** more than $\pm 1\%$ | 6 | |
| Crawling peg | The **central rate** can be **adjusted** | 5 | |
| Crawling band | There are **fluctuation margin**, the **central rate** can be **adjusted** | 6 | |
| Managed floating with no pre-announced path for the exchange rate | **Active** intervention to influence the movement of exchange rate; government has an implicit **target level** for exchange rate | 27 | Large |
| Independently floating | **Few** interventions; no target level for exchange rate | 50 | |

Because pegged arrangement builds a connection between the local currency of a country and the foreign currencies, there exists a web of exchange arrangements among all countries in the world, namely world exchange arrangements web (WEAW). To uncover the topology of this web, we will first build a bipartite network, in which the nodes are categorized into two groups: country nodes and currency nodes, and the edge between every pair of a country node and a currency node represents the exchange arrangement employed by this country. A country employs pegged arrangement will connect those currencies its currency is pegged to. Such a country will also connect to its own local currency. The country with floating arrangement will only connect to its own local currency. Figure 1 is an illustration of a fraction of the WEAW.

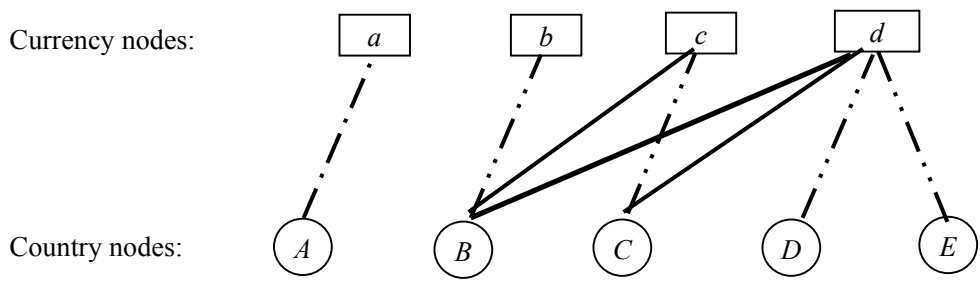

Figure 1. A fraction of the WEAW.

In Fig.1, country *A* only connects to its local currency *a*, and *A* does not peg to any other foreign currencies while its local currency is not pegged by any other countries neither. Some developed countries such as Canada, Japan, Norway belong to this case. Countries *D* and *E* both connect to the same currency *d*, which represents the case of a currency union such as the European Currency Union, and the case of countries that employ exchange arrangement with no separate legal tender, such as those countries holding Eastern Caribbean Dollar and USD as their domestic currencies. Country *B*, representing almost all developing countries, not only connects to its own currency *b*, but also pegs to some other foreign currencies such as *c*. And country *C* is the case of Australia, South African etc., which peg to other currencies while whose local currencies are pegged by other countries. A dashed line is a connection between a country and its own local currency, and a solid line is a connection between a country and its pegged currency. With the statistical data of IMF [21] and the research results of Kawai [22], the WEAW totally has 186 country nodes and 152 currency nodes, and 400 connections among all these country nodes and currency nodes.

## III. Complex topology of the WEAW

One of the most important topological properties of a network is the degree distribution, $P(k)$, which quantifies the probability of a randomly chosen node to have exactly $k$ connections to some other nodes. An alternative measure of $P(k)$ is $P_c(k) \equiv \sum_{k'=k} P(k')$, the cumulative degree distribution. We summed the number of connections (countries), $k$, of each currency, and obtained Fig. 2, where the solid line is a power-law distribution, $P_c(k) = \frac{0.3}{k} \propto k^{-\gamma}$, with $\gamma = 1$.

Hierarchy is generally analyzed by the clustering coefficient and the degree-degree correlation. In the present bipartite network, there are two types of nodes, and no triangles and polygons exist in it. Hence, we cannot calculate the clustering coefficients for this bipartite network. To further study the hierarchical structure of the WEAW, we consider the WEAW in its currency section as a pure currency-node network, as shown in Fig. 3, where a triangle can exist as illustrated.

It can be found that from Fig. 3 that there is an isolated node *a* in the currency section of the WEAW. That is because such a currency is not pegged by any other countries' currencies, and, at the same time, its home country *A* employs the floating arrangement therefore it does not peg to other currency. Such instances include the currencies of 11 countries such as Canada, Denmark, Greece, Iceland, Norway, Sweden, Switzerland, etc. The rebuilt WEAW has 141 currency nodes and 197 connections, whose cumulative degree distribution is log-log plotted in Fig. 4. The solid line is a power-law distribution, $P_c(k) = \frac{0.6}{k^{1.3}} \propto k^{-\gamma}$, with $\gamma = 1.3$. Clearly, both Fig. 2 and Fig. 4 show the scale-free feature of the WEAW.

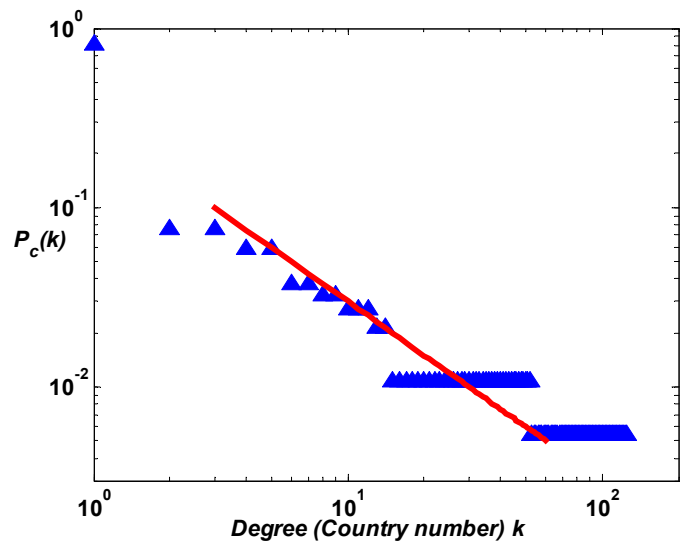

Figure 2. Cumulative degree distribution $P_c(k)$ of the WEAW with 152 currency nodes, 186 country nodes, and 400 connections. The solid line is a power-law $P_c(k) = \dfrac{0.3}{k} \propto k^{-\gamma}$ with $\gamma = 1$. The degree of a currency node means the number of countries connecting to it.

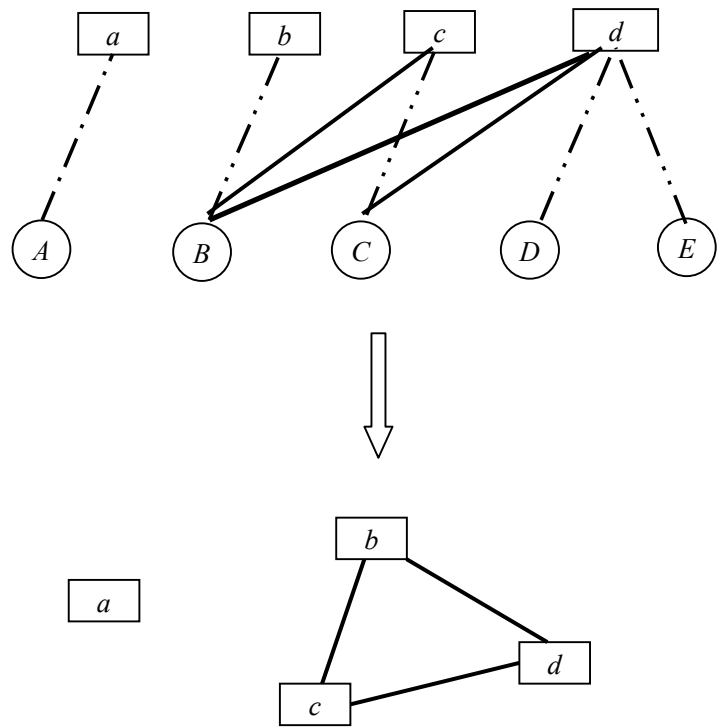

Figure 3. The currency section of the WEAW, where a triangle can be found.

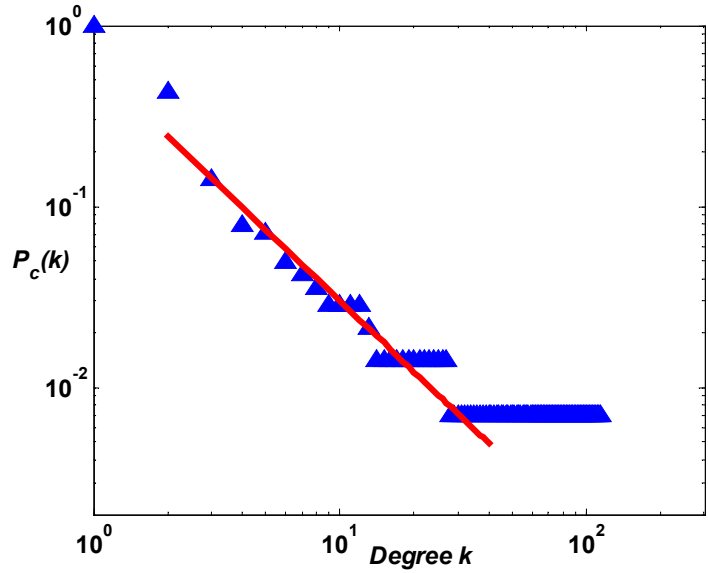

Figure 4. Cumulative degree distribution $P_c(k)$ of the currency section of the WEAW with 151 currency nodes and 197 connections. The solid line is a power-law form $P_c(k) = \dfrac{0.6}{k^{1.3}} \propto k^{-\gamma}$, with $\gamma = 1.3$.

The clustering coefficient of node $i$, of degree $k_i$, is defined as $c_i \equiv \dfrac{2n_i}{k_i(k_i - 1)}$, where $n_i$ is the number of neighbors of node $i$, which are interconnected. Generally, if hierarchy is not presented in a network, $c_i$ is randomly distributed. Very often, $c_i$ exhibits a highly nontrivial behavior with a power-law decay as a function of the node degree $k$, signifying a hierarchy in which low-degree nodes belong to well interconnected communities (with a high clustering coefficient), while hubs connect many nodes that are not directly connected (with a small clustering coefficient) [6, 24]. Based on the available real data, we computed the clustering coefficients of the rebuilt WEAW, and found that, although there are not many clusters in this web, $c_k$ have a declining trend on the node degree $k$. More precisely, those nodes (currencies) with low degrees, such as Nepal Rupee, Nigeria Naira, Botswana Pula, etc., have larger clustering coefficients than those nodes with high degrees, such as Norway Krone, South African Rand, and USD. Furthermore, those currency hubs like USD, ECU, Great Britain Pound (GBP) are not directly connected, which leads to their very small clustering coefficients (USD: 0.0008, ECU and GBP: 0). The averaged clustering coefficient of the whole WEAW is $C = 0.0392$, which is very small but still about 2 times larger as compared with the value 0.02 corresponding to a random network of the same size with the same amount of connections. The average path length of the rebuilt WEAW is the average of the shortest distances between all pairs of currency nodes, and is calculated as $<d>$ =2.43. Compared to the same size of random networks, the WEAW shows a larger clustering coefficient and a similarly small average path length, which indicates this hierarchical web is also a small world.

Another reflection of the web hierarchy is on the degree-degree correlation through the conditional probability $P(k|k')$, which is the probability of a node of degree $k'$ to be linked to a node of degree $k$. To overcome the difficulty in calculating the conditional probability, the correlation is generally quantified by the average nearest-neighbors degree (ANND), which is defined as $<k_{nn}(k)> \equiv \sum_{k'} k' P(k'|k)$.

If a network is uncorrelated, its ANND is independent of $k$. However, it has been discovered that almost all real-life networks, such as the Internet and WTW, show degree-degree correlation [5-6, 16, 24]. The correlation can be assortative or disassortative, depending on the ANND is increasingly or decreasingly dependent on the degree $k$. Figure 5 is the ANND of the rebuilt WEAW, with a clear deceasing dependency on the node's degree. The dashed line and the solid line are both power-law distributions, $<k_{nn}(k)> \propto k^{-v_k}$, with $v_k = 0.9$ and 0.3, respectively.

Therefore, the WEAW is a disassortative network, where highly connected currency nodes trend to connect to poorly connected currency nodes. Furthermore, the web not only is disassortative, it also exhibits two levels of disassortativity: one applies to the majority of nodes with low degrees, the other applies to those with high degrees, to which USD, ECU, GBP, and Japanese Yen belong. Therefore, it is revealed a hierarchical and correlated architecture in the WEAW, and those relatively highly inter-connected currencies that belong to influential areas connect to other influential areas through the hub: USD.

## IV. Conclusions and discussions

In this paper, we have constructed a world exchange arrangements web, the WEAW, and have pointed out that such an international exchange arrangements network falls into the scale-free and small-world networks categories. With the observation on the currency section, this WEAW has been found to be a disassortative network with a hierarchical structure.

As summarized in Table 2, the WTW [14, 16], business network [17], urban economic network [18], and stock price network [19] have also been shown to be scale-free networks. In all these studied economic networks, only the WEAW and the WTW have now been fully discussed on their topological features, and there are indeed some essential differences between them. As pointed out in [16, 23], those low-degree countries in the WTW do not undertake the preferential attachment mechanism, where part of degree distribution is very close to that of an ER random network. However, such a local-world phenomenon has not been found to exist in the WEAW, which means that the scale-free feature is more prominent in the currency exchange market than in the trade web.

It should be noted again that the topology of the WEAW does not have significant change in a long period since the collapse of the Bretton Woods System. Because the scale-free featured web of the international exchange arrangements is almost fixed, the topological influence from the WEAW on the currency crises and economic deflation spreading should cause more attention globally. For example, a viewpoint of stabilizing the world currency is to keep USD, EUR, and JPY stable [20], which are three biggest currency nodes in the WEAW.

Very recently, the phenomena of transition to chaos have been observed in complex dynamical networks [25], which are found to be much easier to be achieved in scale-free networks. As competitions have been proved to accelerate the chaotic butterfly attractors in exchange markets [26], it deserves to further investigate the relation between the discovered scale-free competitive WEAW and the chaotic exchange rate behaviors [27-28].

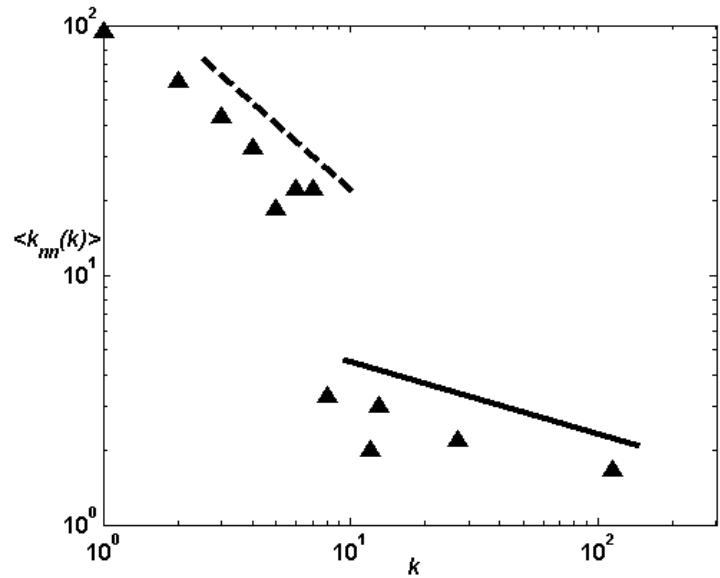

Figure 5. Average nearest-neighbors degree (ANND) of currency nodes as a function of their degrees in the currency section of the WEAW. The dashed line and the solid line are both power-law distributions, $<k_{nn}(k)> \propto k^{-v_k}$, with $v_k = 0.9$ and 0.3, respectively.

**Table 2: Summary of the topological properties of some economic networks**

|  | $n$ | $<k>$ | $<d>$ | $C$ | $\gamma$ | $\omega$ | $v$ |
|---|---|---|---|---|---|---|---|
| WEAW (currency section) | 141 | 2.8 | 2.43 | 0.04 | 1.3 | - | 0.3-0.9 |
| WTW [16] | 179 | 43 | 1.8 | 0.65 | 1.6 | 0.7 | 0.5 |
| WTW [14] | 188 | 66~67.4 | - | - | 0.6~1.6 | - | - |

| Business network [17] | 476 (bank) 13,592 (companies) | - | 1.22[*] | 0.88[*] | 1.4 | - | - |
| Urban economics network [18] | 2.9 million | - | - | - | 2.5~2.8 | - | - |
| Stock price network [19] | 500 | - | - | - | 1.8 | - | - |

[*] The values are calculated in a network of 8 banks and 2388 companies in [17].

## Acknowledgement

The authors are grateful to the anonymous reviewers for their valuable comments and suggestions, which have led to a better presentation of this paper.

## References

[1] R. Albert, A.L. Barabási, Reviews of Modern Physics, **74,** 47, (2002).
[2] D.J. Watts, S.H. Strogatz, Nature, **393**, 440, (1998).
[3] D.J. Watts, Small Worlds, Princeton University Press, RI, USA, (1999).
[4] M.E.J. Newman, SIAM Review, **45**, 167, (2003).
[5] R. Pastor-Satorras, A. Vázquez A., A. Vespignani, Physical Review Letter, **87**, 258701, (2001).
[6] A. Vázquez, R. Pastor-Satorras, A. Vespignani, Phys.Rev. E. **65**, 066130, (2002).
[7] X.F. Wang, International Journal of Bifurcation and Chaos, **12**, 885, (2002).
[8] X.F. Wang, G. Chen, IEEE Circuits and Systems Magazine, **3**, 6, (2003).
[9] P. Erdös, A. Rényi, Publicationes Mathematicae, **6**, 290, (1959).
[10] P. Erdös, A. Rényi, Publications of the Mathematical Institute of the Hungarian Academy of Science, **5**, 17, (1960).
[11] J.E. Stiglitz, Globalization and Its Discontents, The Penguin Press, London, (2002).
[12] Y.Y. Jin, The Stability, Efficiency Faiths and the Regional Exchange Rate Cooperation Mechanism of East Asia (in Chinese), Shanghai University of Finance and Economics Press: Shanghai, (2004).
[13] A. Kirman, Economic networks, in Handbook of Graphs and Networks: From the Genome to the Internet, Bornholdt S. and Schuster H.G. (Eds.), WILEY-VCH GmbH & Co. KgaA: Weiheim, (2003).
[14] X. Li, Y.Y. Jin, G. Chen, Physica A, **328**, 287, (2003).
[15] X. Li, G. Chen, IEEE Transactions on Circuits and Systems-I, **50**, 1381, (2003).
[16] M.A. Serrano, M. Boguñá, Physical Review E, **68**, 015101, (2003).
[17] W. Souma, Y. Fujiwara, H. Aoyama, Physica A, **324**, 396, (2003).
[18] C. Andersson, A. Hellervik, K. Lindgren, A. Hagson, J. Tornberg, eprint: Cond-mat/0303535.
[19] H. Kim, Y. Lee, I. Kim, B. Kahng, eprint: Cond-mat/0107449.
[20] M. Mussa, P. Masson, A. Swoboda, E. Jadresic, P. Mauro, A. Berg, Exchange Rate Regimes in an Increasing Integrated World Economy, Washington D.C.:

International Monetary Fund, (2002).
[21] International Monetary Fund, Annual report on exchange arrangements and exchange restrictions, Washington D.C.: International Monetary Fund, (2000).
[22] M. Kawai, Recommending a currency basket system for emerging east Asia, in Exchange Rate Regimes and Macroeconomic Stability, Ho L. and Yuen C. (eds.), Kluwer Academic Publisher: Boston, (2003).
[23] X. Li, G. Chen, Physica A, **328**, 274, (2003).
[24] A. Barrat, M. Barthelemy, R. Pastor-Satorras, A. Vespignani A., e-print: cond-mat/0311416.
[25] X. Li, G. Chen, K.T. Ko, Physica A, **338**, 367, (2004).
[26] Y. Y. Jin, Competitions hatch butterfly attractors in foreign exchange markets, preprint, (2004).
[27] P. De Grauwe, H. Dewachter, M. Embrechts, Exchange Rate Theory: Chaotic Models of Foreign Exchange Markets, Oxford and Cambridge, MA: Blackwell, (1993).
[28] S. Da Silva, Open Economies Review, **12**, 281, (2001).